\documentclass[preprint,preprintnumbers,showpacs,aps,prd,amssymb]{revtex4-1}

\usepackage{graphicx}
\usepackage{bm}
\usepackage{amsmath}
\def\calH{{\cal H}}
\def\calO{{\cal O}}
\def\calU{{\cal U}}

\def\dU{d_{\calU}}

\def\Bbar{{\bar B}}
\def\Bs{B_s}
\def\Bsbar{{\bar B}_s}
\def\bbar{{\bar b}}
\def\hbar{{\bar h}}

\def\qbar{{\bar q}}

\def\nn{\nonumber}

\begin{document}
\title{Unparticle contributions to $\Bs$-$\Bsbar$ mixing}
\author{Jong-Phil Lee}
\email{jplee@kias.re.kr}
\affiliation{Department of Physics and IPAP, Yonsei University, Seoul 120-749, Korea}
\affiliation{Division of Quantum Phases $\&$ Devices, School of Physics, Konkuk University, Seoul 143-701, Korea}

\begin{abstract}
The unparticle effects on the $\Bs$-$\Bsbar$ mixing is revisited.
Taking into account the unitarity constraints on the unparticle operators, we find that the contribution of the vector unparticle is
very suppressed compared to that of the scalar unparticle.
This is due to the fact that the lower bound of the scaling dimension of the vector-unparticle operator is larger.
It is also shown that the mixing phase from the scalar unparticle is negative, and unparticles can produce large mixing phase.
\end{abstract}

\maketitle
%%%%%%%%%%%%%%%%%%%%%%%%%%%%%%%%%%%%%%%%%%%%%%%%%%%%%%%%%%%%%%%%%%%%%%%%%%%%%%%%
%\section{}
%%%%%%%%%%%%%%%%%%%%%%%%%%%%%%%%%%%%%%%%%%%%%%%%%%%%%%%%%%%%%%%%%%%%%%%%%%%%%%%%
A few years ago Georgi proposed a totally different type of new physics called ``unparticles'' \cite{Georgi}.
In this scenario, there is a scale-invariant hidden sector which couples to the
SM particles very weakly at high energy scale $\Lambda_\calU$.
When seen at low energy, the hidden sector behaves in different ways from ordinary particles, hence dubbed as 
{\em unparticles}.
In a word, unparticles behave like a fractional number of particles.
\par
We have many reasons and clues to conclude that the standard model (SM) of particle physics is only
an effective theory at low energy, and there must be some new physics behind it.
Many kinds of new physics --- supersymmetry or extra dimensions, etc. --- 
involve some new sets of {\em particles},
thus the discovery of the unparticle would be one of the most spectacular phenomena ever seen.
With the reoperation and the first high-energy collision of the Large Hadron Collider (LHC) at CERN very recently,
we anticipate unparticles as well as other new physics signals to be seen sooner or later.
\par
Suppose that at some high energy $\sim M_\calU$, 
there is a ultraviolet (UV) theory in the hidden sector with the infrared (IR)-stable fixed point.
The interaction between the UV theory and the SM sector can be described by an effective theory formalism.
Below $M_\calU$, a UV operator $\calO_{\rm UV}$ interacts with an SM operator $\calO_{\rm SM}$ through 
$\calO_{\rm SM}\calO_{\rm UV}/M_\calU^{d_{\rm SM}+d_{\rm UV}-4}$.
Here $d_{\rm UV(SM)}$ is the scaling dimension of $\calO_{\rm UV(SM)}$.
The renormalization flow enables one to go down along the scale, 
until a new scale $\Lambda_\calU$ is met.
It appears through the dimensional transmutation where the scale invariance emerges.
Below $\Lambda_\calU$ the theory is matched onto the above interaction with
the new unparticle operator $\calO_\calU$ as
\begin{equation}
C_\calU\frac{\Lambda_\calU^{d_{\rm UV}-\dU}}{M_\calU^{d_{SM}+d_{\rm UV}-4}}
\calO_{SM}\calO_{\calU}~,
\end{equation}
where $\dU$ is the scaling dimension of $\calO_\calU$ and $C_\calU$ is the
matching coefficient.
Because of the scale invariance, $\dU$ doesn't have to be integers.
This unusual behavior of unparticles is reflected on the phase space of $\calO_\calU$.
To see it, consider the spectral function of the unparticle which is given by
the two-point function of $\calO_\calU$:
\begin{eqnarray}
\rho_\calU(P^2)&=&\int d^4x~e^{iP\cdot x}
\langle 0|\calO_\calU(x)\calO_\calU^\dagger(0)|0\rangle\nn\\
&=&
A_{\dU}\theta(P^0)\theta(P^2)(P^2)^{\dU-2}~,
\label{rhoU}
\end{eqnarray}
where
\begin{equation}
A_{\dU}=\frac{16\pi^2\sqrt{\pi}}{(2\pi)^{2\dU}}
\frac{\Gamma(\dU+\frac{1}{2})}{\Gamma(\dU-1)\Gamma(2\dU)}~,
\end{equation}
is the normalization factor.
The corresponding phase space is
\begin{equation}
d\Phi_\calU(P)=\rho_\calU(P^2)\frac{d^4P}{(2\pi)^4}
=A_{\dU}\theta(P^0)\theta(P^2)(P^2)^{\dU-2}\frac{d^4P}{(2\pi)^4}~.
\end{equation}
Since $\dU$ is not constrained to be integers, $d\Phi_\calU$  looks like a phase space for a fractional
number of particles. 
\par
After Georgi, there have been a lot of researches on unparticles \cite{Kingman, U}.
Among them are the unparticle effects on $B$-physics and meson mixing \cite{Geng,Li,Mohanta1,Lenz,Chen,Mohanta2,Parry}.
Especially, the $\Bs$-$\Bsbar$ mixing has much attention after the first observation by CDF and D0 \cite{CDF}.
Recently, the D0 collaboration announced the evidence for the charge asymmetry of the like-sign dimuon events \cite{D0}.
For more discussions about $\Bs$-$\Bsbar$ mixing, see \cite{kylee,Lenz2} and references therein.
\par
For simplicity we only consider the left-handed currents coupled to scalar($\calO_\calU$) and vector($\calO_\calU^\mu$)
unparticles as follows:
\begin{equation}
 \frac{c_S}{\Lambda_\calU^{\dU}}\qbar'\gamma_\mu(1-\gamma_5)q~\partial^\mu\calO_\calU
+\frac{c_V}{\Lambda_\calU^{\dU-1}}\qbar'\gamma_\mu(1-\gamma_5)q~\calO_\calU^\mu~,
\label{L}
\end{equation}
where $c_{S,V}$ are dimensionless coefficients.
We assume that $c_{S,V}$ are real numbers.
The above interactions provide flavor-changing neutral currents at tree level, which contribute to the $B_s$-$\Bbar_s$ mixing.
The propagators of scalar and vector unparticles are given by \cite{Georgi,GIR}
\begin{equation}
 \int d^4x~e^{iPx}\langle 0|T\calO_\calU(x)\calO_\calU(0)|0\rangle
=\frac{iA_{\dU}}{2\sin\dU\pi}\frac{e^{-i\phi_\calU}}{(P^2+i\epsilon)^{2-\dU}}~,
\label{SP}
\end{equation}
and
\begin{equation}
 \int d^4x~e^{iPx}\langle 0|T\calO_\calU^\mu(x)\calO_\calU^\nu(0)|0\rangle
=\frac{iA_{\dU}}{2\sin\dU\pi}\frac{e^{-i\phi_\calU}}{(P^2+i\epsilon)^{2-\dU}}
   \left[-g^{\mu\nu}+\frac{2(\dU-2)}{\dU-1}\frac{P^\mu P^\nu}{P^2}\right]~,
\label{VP}
\end{equation}
respectively, and $\phi_\calU=(\dU-2)\pi$.
Note that the relative size of the coefficients of $-g^{\mu\nu}$ and $P^\mu P^\nu/P^2$ in Eq.\ ({\ref{VP}) is not unity, 
but some function of $\dU$ \cite{GIR}.
It is due to the unitarity constraints.
This point is not reflected in the literature.
Another point which is erroneously used so far is that the scaling dimension $\dU$ is commonly used for 
$\calO_\calU$ and $\calO_\calU^\mu$.
Obviously this is not true; in general they can be independent variables.
Furthermore, \cite{GIR} has shown that the scalar-unparticle dimension has a lower bound $\dU^S\ge 1$
while for the vector-unparticle dimension, $\dU^V\ge 3$ from unitarity \cite{GIR}.
Thus in what follows, we will distinguish $\dU^S\equiv d_S$ and $\dU^V=d_V$.
As will be seen later, the unitarity bound for $d_V$ has a significant meaning for the $B_s$-$\Bbar_s$ mixing.
\par
In general, the $\Bs$-$\Bsbar$ mixing is parametrized by the quantity $M_{12}^s$ defined by 
\begin{equation}
 2M_{\Bs}M_{12}^s=\langle\Bsbar^0|\calH_{eff}^{\Delta B=2}|\Bs^0\rangle~,
\end{equation}
where $\calH_{eff}^{\Delta B=2}$ is the effective Hamiltonian for the $\Delta B=2$ transitions.
The SM contribution to $M_{12}^s$ is given by the box diagrams, resulting in
\begin{equation}
 M_{12}^s=\frac{G_F^2M_W^2}{12\pi^2}(V_{ts}^*V_{tb})^2 M_{\Bs}B_{\Bs}f_{\Bs}^2{\hat\eta}_{\Bs}S_0(x_t)~,
\end{equation}
where $S_0(x_t\equiv m_t^2/M_W^2)$ is the Inami-Lim function \cite{Inami} and ${\hat\eta}_{\Bs}$ is the 
QCD correction factor.
The mass difference $\Delta M_s$ is then $\Delta M_s=2|M_{12}^s|$, and the experimentally measured value is 
\cite{CDF}
\begin{equation}
 \Delta M_s^{\rm exp}=17.77\pm0.12~{\rm ps}^{-1}~.
\end{equation}
\par
If there is a new interaction of Eq.\ (\ref{L}), it contributes to the $\Bs$-$\Bsbar$ mixing through the $s$- and $t$-channels
at tree level.
Explicitly, one gets
\begin{eqnarray}
 M_{12}^\calU&=&
\frac{A_{d_S}e^{-i\phi_{\calU_S}}}{8\sin d_S\pi}\left(\frac{f_{\Bs}^2}{M_{\Bs}}\right)c_S^2
\left(\frac{M_{\Bs}^2}{\Lambda_\calU^2}\right)^{d_S}\frac{m_b^2}{M_{\Bs}^2}\frac{5}{3}RB_2\nn\\
&&
+\frac{A_{d_V}e^{-i\phi_{\calU_V}}}{8\sin d_V\pi}\left(\frac{f_{\Bs}^2}{M_{\Bs}}\right)c_V^2
\left(\frac{M_{\Bs}^2}{\Lambda_\calU^2}\right)^{d_V-1}
\left[-\frac{8}{3}B_1+\frac{2(d_V-2)}{d_V-1}\frac{m_b^2}{M_{\Bs}^2}\frac{5}{3}RB_2\right]~,
\label{M12U}
\end{eqnarray}
where 
\begin{equation}
 R\equiv \left(\frac{M_{\Bs}}{m_b+m_s}\right)^2~.
\end{equation}
Here $B_{1,2}$ are the bag parameters for the relevant operators as follows:
\begin{eqnarray}
 \langle\Bsbar|Q_1|\Bs\rangle&=&\frac{8}{3}M_{\Bs}^2f_{\Bs}^2B_1~,\\
 \langle\Bsbar|Q_2|\Bs\rangle&=&-\frac{5}{3}M_{\Bs}^2f_{\Bs}^2RB_2~,
\end{eqnarray}
where
\begin{eqnarray}
 Q_1&=&\bbar_\alpha\gamma_\mu(1-\gamma_5)s_\alpha\bbar_\beta\gamma^\mu(1-\gamma_5)s_\beta~,\\
 Q_2&=&\bbar_\alpha(1-\gamma_5)s_\alpha\bbar_\beta(1-\gamma_5)s_\beta~.
\end{eqnarray}
\par
The new physics effects on $\Bs$-$\Bsbar$ mixing can be nicely encoded in the following manner \cite{Nierste}:
\begin{equation}
 M_{12}=M_{12}^{\rm SM}+M_{12}^\calU\equiv M_{12}^{\rm SM}\cdot \Delta~.
\end{equation}
The phase of $M_{12}$ is 
\begin{equation}
 \phi_s=\phi_s^{\rm SM}+\phi_s^\Delta~,
\end{equation}
where $\Delta=|\Delta|~e^{i\phi_s^\Delta}$.
With the help of Eq.\ (\ref{M12U}), one can easily obtain (for simplicity we put $m_b=M_{\Bs}$, and $B_{1,2}=R=1$)
\begin{eqnarray}
 \Delta&=&1+\frac{M_{12}^\calU}{M_{12}^{\rm SM}}\nn\\
&=&\Big[1+c_S^2 f_S(d_S)\cot d_S\pi+c_V^2 f_V(d_V)\cot d_V\pi\Big]
     -i\Big[c_S^2 f_S(d_S)+c_V^2 f_V(d_V)\Big]~,
\label{Delta}
\end{eqnarray}
where
\begin{eqnarray}
 f_S(d_S)&\equiv&\frac{1}{M_{12}^{\rm SM}}\left(\frac{f_{\Bs}^2}{M_{\Bs}}\right)\frac{2\pi^{5/2}}{(2\pi)^{2d_S}}
    \frac{\Gamma(d_S+\frac{1}{2})}{\Gamma(d_S-1)\Gamma(2d_S)}\left(\frac{M_{\Bs}^2}{\Lambda_\calU^2}\right)^{d_S}
    \frac{5}{3}~,\\
  f_V(d_V)&\equiv&\frac{1}{M_{12}^{\rm SM}}\left(\frac{f_{\Bs}^2}{M_{\Bs}}\right)\frac{2\pi^{5/2}}{(2\pi)^{2d_V}}
    \frac{\Gamma(d_V+\frac{1}{2})}{\Gamma(d_V-1)\Gamma(2d_V)}\left(\frac{M_{\Bs}^2}{\Lambda_\calU^2}\right)^{d_V-1}
    \frac{2(d_V-6)}{3(d_V-1)}~.
\label{fSV}
\end{eqnarray}
Note that the power of $M_{\Bs}^2/\Lambda_\calU^2$ is different for $f_S$ and $f_V$.
If $d_S=d_V$, then $f_S$ is suppressed by a factor of $M_{\Bs}^2/\Lambda_\calU^2$,
which amounts to $\sim 3\times 10^{-5}$ for $\Lambda_\calU=1$ TeV.
But if we consider the unitarity constraints, $d_S\ge 1$ and $d_V\ge 3$.
For simplicity we may set $d_S=1+\epsilon$, $d_V=3+\epsilon$.
In this case, on the contrary to the previous estimation, $f_V$ is much more suppressed by the factor of
\begin{equation}
\sim \frac{1}{(2\pi)^4}\left (\frac{M_{\Bs}^2}{\Lambda_\calU^2}\right)=1.8\times 10^{-8}~,
\label{supp}
\end{equation}
for $\Lambda_\calU=1~{\rm TeV}$.
\par
The experimental values for $\Delta M_s$ and $\phi_s^\Delta$ constrains the new physics parameters.
Figure \ref{cSe} shows the allowed region of $c_S$ and $\epsilon$ when $c_V=0$.
We use one of the latest value of $\phi_s=-0.79\pm 0.24$ \cite{Bauer} which fits the new D0 anomalous dimuon asymmetry,
and $\phi_s^{\rm SM}=(4.7^{+3.5}_{-3.1})\times 10^{-3}$ \cite{Lenz2}.
Even in the case of $c_V\ne 0$, the effect of $c_V\sim\calO(1)$ is negligible 
because of the suppression by Eq.\  (\ref{supp}).
%------------------------- FIGURE 1 -------------------------------------------
\begin{figure}
\includegraphics{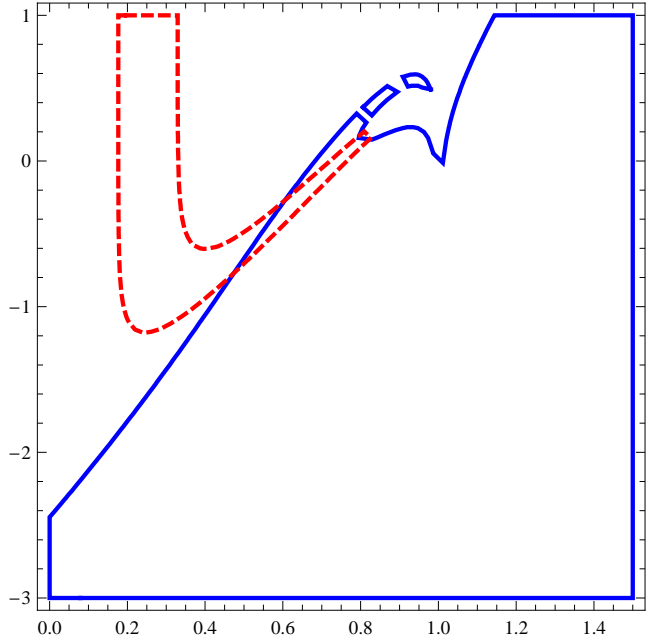}
\caption{Allowed region of $c_S$ (vertical, in log scale) and $\epsilon$ (horizon) 
from experimentally measured $\Delta M_s$ and $\phi_s^\Delta$ for $c_V=0$. 
Blue region (solid line) is from $\Delta M_s^{\rm exp}$ (1-$\sigma$) 
while red one (dashed line) from $\phi_s^\Delta$ (1-$\sigma$).}
\label{cSe}
\end{figure}
%------------------------------------------------------------------------------
If we switch off $c_S$ and turn on $c_V$, we have no overlaps for $\Delta M_s^{\rm exp}$ and $\phi_s^\Delta$,
at least for moderate ranges of $c_V$ and $\epsilon$.
In other words, the coupling $c_V$ must be enormous to compensate the kinematic suppression (\ref{supp}).
\par
The expression Eq.\ (\ref{Delta}) also has important meanings for the phase, $\phi_s^\Delta$.
The imaginary part of $\Delta$ is 
\begin{equation}
c_S^2f_S+c_V^2f_V
=-|\Delta|~\sin\phi_s^\Delta 
=-\left(\frac{\Delta M_s}{\Delta M_s^{\rm SM}}\right)\sin\phi_s^\Delta~.
\label{sinphi}
\end{equation}
Since $f_V$ is highly suppressed, the left-hand-side remains positive (for moderate values of $c_V$) 
and thus $-\pi<\phi_s^\Delta<0$.
Note that our $M_{12}^\calU$ is the same as that of \cite{Parry}, 
and different from \cite{Lenz} by a factor of $(i/2)$.
For this reason, the $\cot (d_{S,V}\pi)$ term enters the imaginary part of $\Delta$ in \cite{Lenz}
and the phase can have both positive and negative values with the variation of $d_{S,V}$.
In our calculation this is not true.
Figure \ref{Dphi} shows $\Delta M_s$ vs $\phi_s^\Delta$ for various values of $c_S$.
%------------------------- FIGURE 2 -------------------------------------------
\begin{figure}
\includegraphics{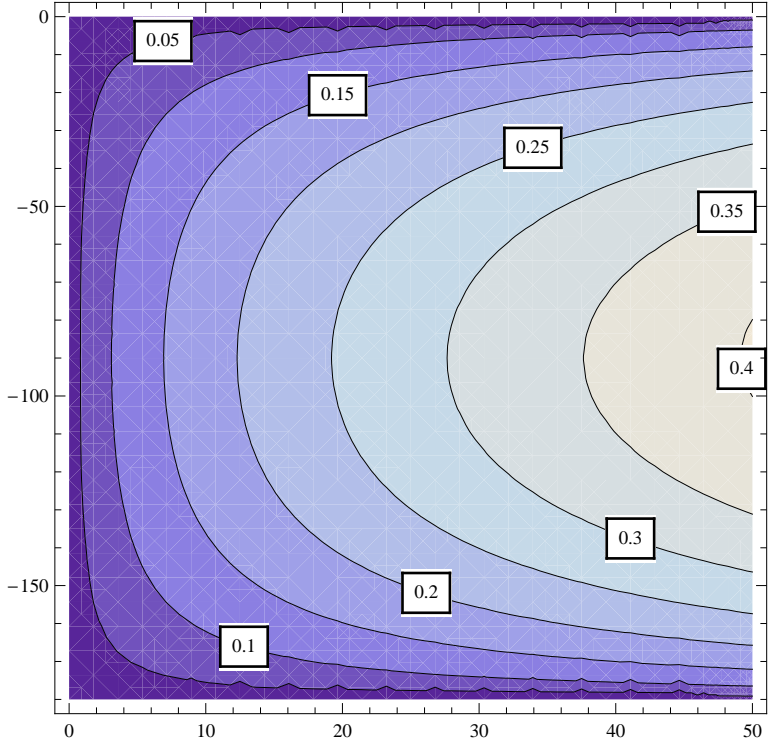}
\caption{Contour plots for $\Delta M_s$ (horizon, in ${\rm ps}^{-1}$) and $\phi_s^\Delta$ (vertical, in degree)
for $\epsilon=0.5$. We put $c_V=0$. The numbers in the boxes are the values of $c_S$.}
\label{Dphi}
\end{figure}
%------------------------------------------------------------------------------
In this Figure, we only consider the scalar contribution.
For $c_V=0$, the imaginary part of $\Delta$ is definitely negative, 
so we expect that the scalar unparticles induce negative $\sin\phi_s^\Delta$.
As one can easily find in the Figure, the scalar unparticle can produce a large phase.
\par
If $\phi_s^\Delta$ turned out to be positive, then one could expect $c_S=0$ and $c_V\ne 0$.
Note that for $\epsilon<3$ the function $f_V$ is negative.
But the suppression is very severe, and the coupling $c_V$ should be of order $\sim\calO(10^8)$.
So in this case one can conclude that the unparticle contributions cannot explain the positive $\phi_s^\Delta$ for 
moderate values of the couplings.
\par
In conclusion, we investigated the unparticle effects on the $\Bs$-$\Bsbar$ mixing.
Scalar and vector unparticles can contribute to the mixing at tree level via $s$- and $t$-channels of the unparticle exchange.
The effects were already studied in the literature, but the previous studies did not consider the unitarity constraints of
\cite{GIR}.
We found that the unitarity constraints play a crucial role in the analysis.
If the scaling dimensions of the unparticle operators are universal as is common in the literature, 
the vector-unparticle contribution is dominant.
But the unitarity condition puts different lower bounds for the dimensions of the unparticle operators.
When we take into account this point, the vector-unparticle contribution is highly suppressed by a factor of 
$\sim\calO(10^{-8})$, compared to the scalar-unparticle contribution (assuming that the couplings are of the same order).
According to \cite{GIR}, the tensor structure of the propagator of the vector unparticles is slightly different from that of the
ordinary vector particles.
But since the vector contribution to the $\Bs$-$\Bsbar$ mixing is negligible, it is very hard to notice the differences.
We also found that the phase $\phi_s^\Delta$ from the scalar unparticle is negative definite.
This is compatible with the current experimental data.
Fortunately, the scalar unparticle can produce large mixing phase.
\par
It might be also interesting to examine the unparticle effects on $\Bs\to J/\psi\phi$ and $\Bs\to\phi\phi$, 
as analyzed in \cite{Mohanta1}.
With the unitarity constraints, the fact that the mixing-induced CP asymmetry for $\Bs\to J/\psi\phi$ 
can be large by unparticles would not be changed, but the contribution would be dominated by scalar unparticles.
And the transition amplitude of $\Bs\to\phi\phi$ from vector unparticles is much more suppressed compared to the result
of \cite{Mohanta1} since the amplitude is proportional to $(m_{\Bs}/\Lambda_\calU)^{2d_V-2}$.
\par
As a final remark, possible new physics effects on the decay matrix element $\Gamma_{12}^s$ have received much attention 
recently after the D0 anomaly.
There have been lots of works considering new physics effects on $\Gamma_{12}^s$.
Since $\Gamma_{12}^s$ is the absorptive part of the effective Hamiltonian and unparticles can be seen as an infinite tower
of massless particles \cite{Stephanov},
one could expect that there is a sizable contribution from unparticles \cite{He}.
Dedicated works to this issue will appear elsewhere.
In the current analysis we simply assumed that new physics contributes only to $M_{12}$.
%%%%%%%%%%%%%%%%%%%%%%%%%%%%%%%%%%%%%%%%%%%%%%%%%%%%%%%%%%%%%%%%%%%%%%%%%%%%%%%%
\begin{acknowledgments}
This work was supported by the Basic Science Research Program through the National Research Foundation of Korea (NRF) 
funded by the Korean Ministry of Education, Science and Technology (2009-0088396).
\end{acknowledgments}
%%%%%%%%%%%%%%%%%%%%%%%%%%%%%%%%%%%%%%%%%%%%%%%%%%%%%%%%%%%%%%%%%%%%%%%%%%%%%%%%
%%%%%%%%%%%%%%%%%%%%%%%%%%%%%%%%%%%%%%%%%%%%%%%%%%%%%%%%%%%%%%%%%%%%%%%%%%%%%%%%

\end{document}